\documentclass{PoS}

\newcommand{\Or}{\textrm{O}}
\newcommand{\U}{\textrm{U}}

\newcommand{\tr}{\textrm{tr}}

\newcommand{\diag}{\textrm{diag}}

\newcommand{\eins}{\leavevmode\hbox{\small1\kern-3.8pt\normalsize1}}

\title{Random Matrix Models for Dirac Operators at finite Lattice Spacing}

\ShortTitle{RMT for Lattice Dirac Operators}

\author{Mario Kieburg\\
        Department of Physics and Astronomy, State University of New York at Stony Brook, NY 11794-3800, USA\\
        E-mail: \email{mario.kieburg@stonybrook.edu}}

\author{Jacobus J. M. Verbaarschot\\
        Department of Physics and Astronomy, State University of New York at Stony Brook, NY 11794-3800, USA\\
        E-mail: \email{jv@chi.physics.sunysb.edu}}

\author{Savvas Zafeiropoulos\\
        Department of Physics and Astronomy, State University of New York at Stony Brook, NY 11794-3800, USA\\
        E-mail: \email{szafeiro@ic.sunysb.edu}}

\abstract{We study discretization effects of the Wilson  and staggered Dirac operator with $N_{\rm c}>2$ using chiral random matrix theory (chRMT). We obtain  analytical results for the joint
 probability density of Wilson-chRMT in
 terms of a determinantal expression over complex pairs of eigenvalues, and real eigenvalues corresponding to eigenvectors of
 positive or negative chirality as well as for the eigenvalue densities. The explicit
 dependence on the lattice spacing can be readily read off from our results which are compared to
 numerical simulations of Wilson-chRMT. For the staggered Dirac operator we have studied random
 matrices modeling the transition from non-degenerate eigenvalues at non-zero lattice spacing to
 degenerate ones in the continuum limit.}

\FullConference{XXIX International Symposium on Lattice Field Theory \\
		 		  July 10 -- 16 2011\\
		 		  Squaw Valley, Lake Tahoe, California}

\begin{document}

\section{Introduction}\label{sec1}

 Chiral RMT is a powerful mathematical tool to calculate
 eigenvalue correlations in the infrared limit of quantum chromodynamics (QCD) and has
 been successfully applied to Dirac spectra for almost two decades \cite{ShuVer93}. In the low energy limit QCD
 exhibits universal behavior that agrees with chRMT with the same symmetries as the Dirac
 operator. In the continuum limit as well as  for finite lattice spacing, $a\neq0$, random
 matrix ensembles have been constructed that reproduce the  eigenvalue correlations of the lattice QCD Dirac operator. 

 Discretization effects of the Wilson Dirac operator were already studied with the help of chiral
 perturbation theory in the $p$-limit \cite{Sharpe:1998xm,BRS04,necco}. Furthermore it was shown
 that the chiral Lagrangian agrees with Wilson chRMT in this limit
 \cite{DSV10,ADSV10b,SplVer11}. Random matrix theory enables us to get results which
 were not accessible before. We derive the joint probability density (jpd) of the non-Hermitian version
 for the random matrix ensemble proposed in Ref.~\cite{DSV10} and present its eigenvalue
 densities in the microscopic limit which have to agree  with the low lying eigenvalues of the Wilson
 Dirac operator. In our representation we are able to distinguish between the eigenvalue densities of
 the complex eigenvalues and of the real eigenvalues with eigenvectors of positive and negative
 chirality.

 Staggered fermions are widely used in lattice simulations mainly because of their
 low computational costs. We study a matrix model that reproduces the
 single trace terms  in the chiral Lagrangian (reduced to staggered fermions in two dimensions) 
incorporating  taste breaking effects at order $a^2$
 \cite{LeeShar}. It is the model proposed in Ref.~\cite{Osb10} for staggered fermions in two dimensions which we expect to be analytically solvable.

 In Sec.~\ref{sec2}, we consider the Wilson-chRMT \cite{DSV10} and study its spectral properties. A
 random matrix ensemble for the staggered fermions is proposed in Sec.~\ref{sec3}.

\section{Wilson fermions}\label{sec2}


\paragraph{Wilson-chRMT.}

The ensemble introduced in Ref.~\cite{DSV10} is
\begin{eqnarray}\label{2.1.1}
 D_{\rm W}&=&\left(\begin{array}{cc} aA & W \\ -W^\dagger & aB\end{array}\right)\quad{\rm 
 distributed\ by}\quad P(D_{\rm W})\propto\exp\left[-\frac{n}{2}(\tr A^2+\tr B^2)-n\tr WW^\dagger\right].
\end{eqnarray}
For $a=0$ this gives the chiral Gaussian unitary ensemble (chGUE) that describes the
 infrared Dirac spectrum of  continuum QCD \cite{ShuVer93}.
 The  complex $n\times (n+\nu)$ matrices $W$ and $W^\dagger$ preserve chiral symmetry, whereas the Hermitian matrices 
$A$ and $B$ break the chiral symmetry
 and are identified with the Wilson term. In the microscopic limit ($n\to\infty$) the rescaled lattice spacing
 $\widehat{a}^2=na^2/2$, the rescaled eigenvalues $\widehat{z}=2nz$ as well as the index of the Dirac
 operator $\nu$ are kept fix. Then, spectral correlations of Wilson-chRMT become universal and agree
 with infrared Wilson chiral perturbation theory ($\chi$PT) \cite{ADSV10b,SplVer11}. The
 volume of space-time $V$ is identified with the matrix dimension $n$.
 
 Let $\widetilde{a}$ be the physical lattice spacing. Then, the relation to the rescaled quantity
 is $\widehat{a}=\sqrt{W_8V}\widetilde{a}$.
 The low energy constants $W_6$ and $W_7$ corresponding to the squares of traces in the action of the
 Goldstone bosons \cite{DSV10,ADSV10b} are suppressed in the large $N_{\rm c}$ limit 
 \cite{Kaiser:2000gs} and will not be
 considered here. We are interested in the quenched case and a more general setting will be considered
 elsewhere.

 Without loss of generality, let $\nu\geq0$. Due to the
 $\gamma_5=\diag(\eins_n,-\eins_{n+\nu})$-Hermiticity of $D_{\rm W}$ ($(\gamma_5D_{\rm
 W})^\dagger=\gamma_5D_{\rm W}$), the eigenvalues are either real or come in complex conjugated pairs.
 There are $\nu$ generic real eigenvalues corresponding to the $\nu$ zero modes at $a=0$. Moreover, $2l$
 ($0\leq l\leq n$) additional real modes may appear when $l$ complex conjugated eigenvalue
 pairs enter the real axis.

 Often the Hermitian version of the Wilson Dirac operator $D_5=\gamma_5D_{\rm W}$ is studied because 
 it simplifies lattice
 simulations \cite{Luscher} as well as RMT calculations \cite{DSV10,ADSV10b,SplVer11,Akemann:2011kj}.
 However, only $D_{\rm W}$ is directly related to the chiral symmetry
 breaking which is our main motivation for studying the non-Hermitian version.

\paragraph{The joint probability distribution.}

 Analogously to Hermitian matrices, the $\gamma_5$ Hermiticity allows us to quasi-diagonalize the matrix
 $D_{\rm W}$ by a non-compact unitary matrix $U\in\U(n,n+\nu)$,
\begin{eqnarray}\label{2.2.1}
 D_{\rm W}=UX^{(l)}U^{-1}\ \textrm{and}\ X^{(l)}=\diag\left[x_1,\left(\begin{array}{cc}  x_2 & y_2 \\
 -y_2 & x_2 \end{array}\right),x_3\right].
\end{eqnarray}
 The real matrices $x_1$, $x_2$, $y_2$ and $x_3$ are diagonal with dimensions $l$, $n-l$, $n-l$ and
 $l+\nu$, respectively. The ensemble decomposes in $n+1$ disjoint sets differing in the
 number of complex conjugate pairs, $n-l$, or equivalently, in the number of real modes, $2l+\nu$. The $l$
 complex pairs are given by $(z_2,z^*_2)=(x_2+\imath y_2,x_2-\imath y_2)$.

 Let $Z=(z_{1{\rm r}},\ldots,z_{n{\rm r}},z_{1{\rm l}},\ldots,z_{n+\nu,{\rm l}})\in\mathbb{C}^{2n+\nu}$
 be the $2n+\nu$ eigenvalues of $D_{\rm W}$ where we ignore which of them are complex or  real.
 The jpd is given by a sum over non-compact coset integrals
\begin{equation}\label{2.2.2}
 p(Z)d[Z]\propto\Delta_{2n+\nu}^2(Z)\sum\limits_{l=0}^n\int_{\mathbb{G}_l}P(UX^{(l)}U^{-1})
 d\mu_{\mathbb{G}_l}(U)d[X^{(l)}]
\end{equation}
 with the Haar measure $d\mu_{\mathbb{G}_l}$ on the coset
 $\mathbb{G}_l=\U(n,n+\nu)/[\U^{n+l+\nu}(1)\times\Or^{n-l}(1,1)]$ and the Vandermonde determinant $\Delta_{2n+\nu}(Z)$. After the integration we are able
 to perform the sum by introducing Dirac delta functions and find \cite{Kieburg:2011uf}
\begin{eqnarray}
 p(Z)&\propto&\Delta_{2n+\nu}(Z)\det\left[\begin{array}{c}  \displaystyle\left\{g_2(z_{a{\rm
 r}},z_{b{\rm l}})\right\}_{ 1\leq a\leq n,\ 1\leq b\leq n+\nu} 
 \\ \displaystyle\left\{z_{b{\rm l}}^{a-1}\sqrt{\frac{n}{2\pi a^2}}\exp\left[-\frac{n}{2a^2}x_{b{\rm
 l}}^2\right]\delta(y_{b{\rm l}})\right\}_{1\leq a\leq \nu,\ 1\leq b\leq n+\nu}
  \end{array}\right],\label{2.2.3}\\
 g_2(z_1,z_2)&=&
 \sqrt{\frac{n^3}{4\pi a^2(1+a^2)}}\frac{z_1^*-z_2^*}{|z_1-z_2|}
 \left[\exp\left[-\frac{n(x_1+x_2)^2}{4a^2}-\frac{n(y_1-y_2)^2}{4}\right]
 \delta^{(2)}(z_1-z_2^*)\right.\nonumber\\
 &+&\left.\frac{1}{2}\exp\left[-\frac{n}{4a^2}(x_1+x_2)^2+\frac{n}{4}(x_1-x_2)^2\right]{\rm
 erfc}\left[\sqrt{n(1+a^2)}\frac{|x_1-x_2|}{2a}\right]\delta(y_1)\delta(y_2)\right],\nonumber\\
 &\equiv& g_{2\,{\rm  c}}(z_1)\delta^{(2)}(z_1-z_2^*)+
 g_{2\,{\rm  r}}(x_1,x_2)\delta(y_1)\delta(y_2).\label{2.2.4}
\end{eqnarray}
 The function ${\rm erfc}$ is the complementary error function and
 $\delta^{(2)}(x+\imath y)=\delta(x)\delta(y)$. The $\gamma_5$-Hermiticity enforces a breaking of the
 permutation group $S(2n+\nu)$ to $S(n)\times  S(n+\nu)$ which is
 reflected in the expression~(\ref{2.2.3}). An expansion in the Dirac delta functions yields all $n+1$
 terms in the sum~(\ref{2.2.2}) corresponding to a fixed number of real modes. The two-point weight
 $g_2$ splits into a weight for the real modes, $g_{2\,{\rm  r}}$, and for the complex conjugated pairs,
 $g_{2\,{\rm  c}}$. Such a structure is already known from the real Ginibre ensemble \cite{SomWie08} and
 its chiral counterpart \cite{Gin}.

\paragraph{The eigenvalue densities.}

 The integration over all eigenvalues except one yields the three eigenvalue densities of real modes,
 $\rho_{\rm r}$ for positive chirality ($\langle\psi|\gamma_5|\psi\rangle>0$), $\rho_{\rm l}$ for
 negative chirality ($\langle\psi|\gamma_5|\psi\rangle<0$), and of complex pairs, $\rho_{\rm c}$, i.e.
\begin{eqnarray}
 \hspace*{-0.2cm}\int p(Z)\prod\limits_{z_j\neq z_{1{\rm r}}}d[z_j]
 =\rho_{\rm r}(x_{1{\rm r}})\delta(y_{1{\rm r}})+ \frac{\rho_{\rm c}(z_{1{\rm r}})}{2},\quad\int
 p(Z)\prod \limits_{z_j\neq z_{1{\rm l}}}d[z_j]
 =\rho_{\rm l}(x_{1{\rm l}})\delta(y_{1{\rm l}})+\frac{\rho_{\rm c}(z_{1{\rm l}})}{2}\label{2.3.2}.
\end{eqnarray}
 In both integrals the determinant~(\ref{2.2.3}) can be expanded either in the first row or in the first column. The densities $\rho_{\rm c}$ and $\rho_{\rm r}$ can be expressed in terms of  the
 $N_{\rm f}=2$ partition function
\begin{eqnarray}
 \hspace*{-0.2cm}\rho_{\rm c}(z)=g_{2{\rm c}}(z)(z-z^*) Z_{N_f=2}^\nu(z,z^*;a),\quad \rho_{\rm
 r}(x)=\int dy g_{2{\rm r}}(x,y)(x-y) Z_{N_f=2}^\nu(x,y;a). \label{2.3.4}
\end{eqnarray}
 The eigenvalue distribution of the left handed modes contains an additional contribution from the $\nu$
 generic real modes which is the chiral distribution $\rho_\chi = \rho_{\rm l} -\rho_{\rm r}$.
 To express $\rho_\chi$ into known partition functions additional rows of some of the determinants
 have to be expanded. For $\nu =1$ and $\nu =2 $ we checked that this result agrees with previously
 derived expressions \cite{DSV10,SplVer11}.

\begin{figure}
 \includegraphics[width=0.49\textwidth]{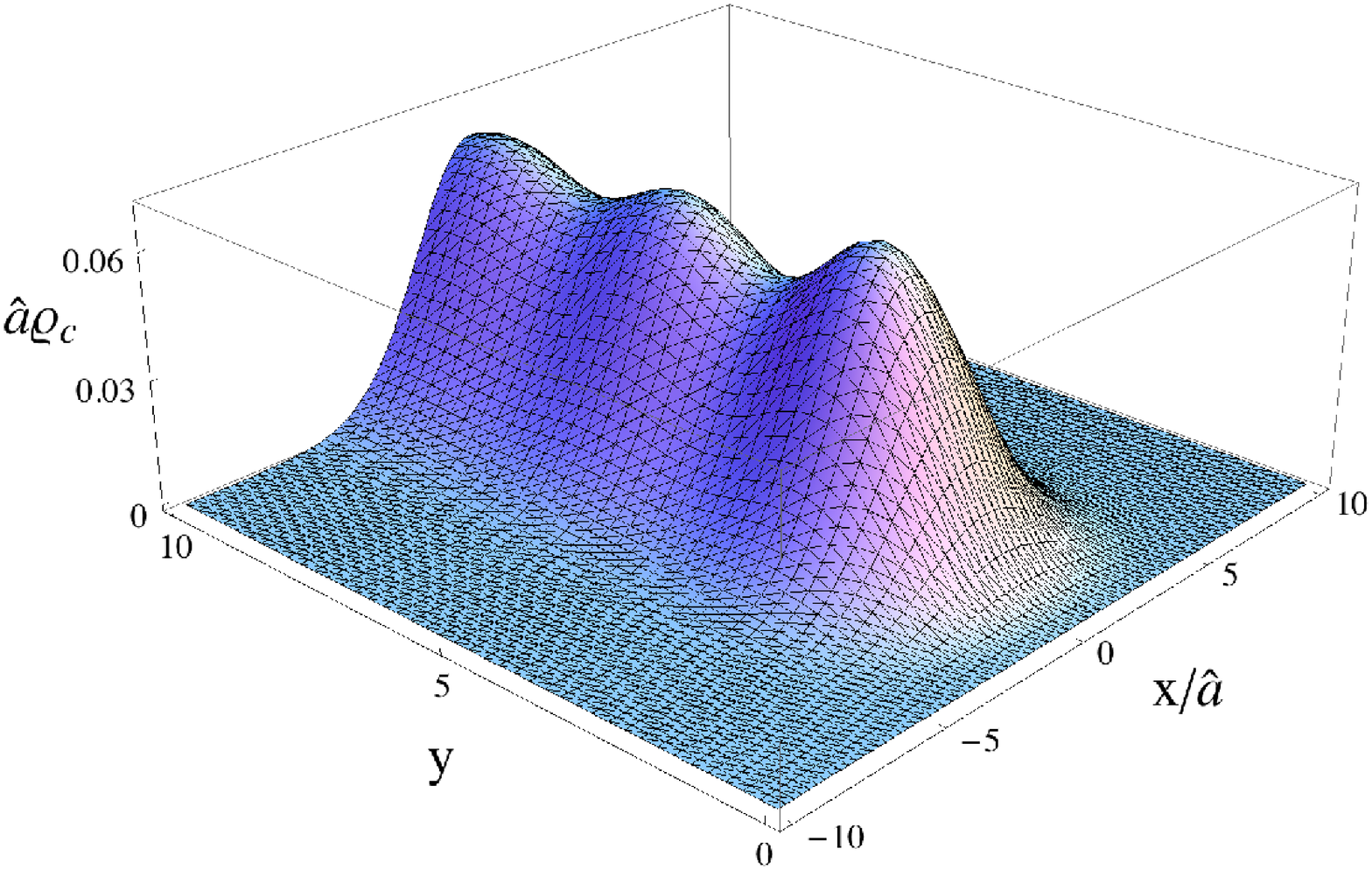}
 \includegraphics[width=0.49\textwidth]{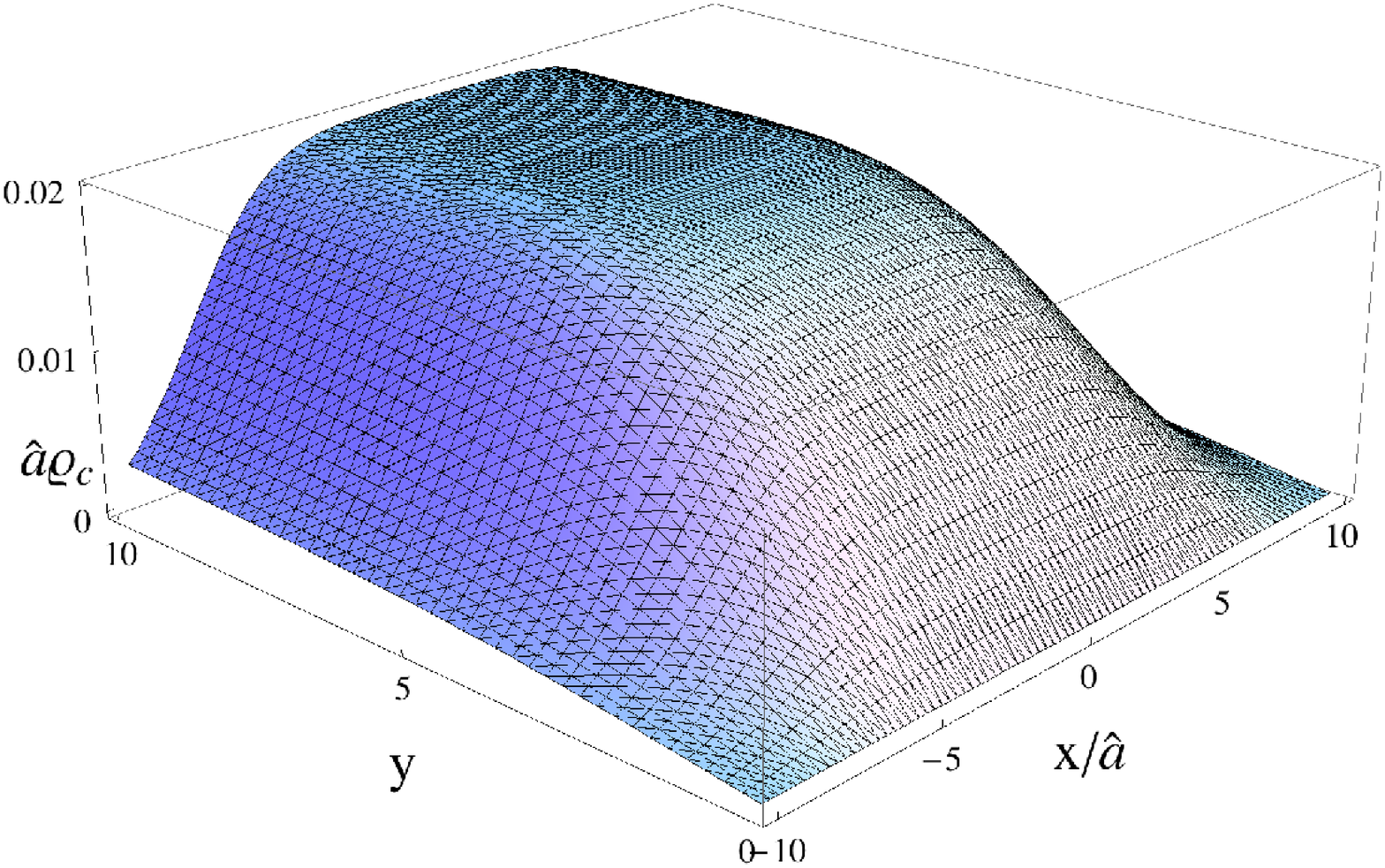}\label{fig1}
 \caption{These 3d-plots show $\rho_{\rm c}$ for small lattice spacing ($\widehat{a}=0.1$,
 left figure) and for a large one ($\widehat{a}=1$, right figure). The index is $\nu=1$. 
Oscillations are well distinguishable for small $\widehat{a}$ whereas they
 disappear for larger lattice spacing. Notice that $\widehat{a}$ in the right figure is one 
 order larger than in the left one. }
\end{figure}

 In the microscopic limit ($n\to\infty$) we obtain \cite{Kieburg:2011uf}
\begin{eqnarray}
 \hspace*{-0.9cm}\rho_{\rm c}\left(\frac{z}{2n}\right)&=&\frac{|y|}{(2\pi)^{5/2}2\widehat{a}}
 \int \exp(-\Delta_1^2-\Delta_2^2)
 {\rm sinc}\left[\frac{y}{2\widehat{a}}(\Delta_1-\Delta_2)\right]\cos\nu(\varphi_1+\varphi_2)
 D\varphi_k,\label{2.3.6}\\
 \hspace*{-0.9cm}\rho_{\rm r}\left(\frac{x}{2n}\right)&=&
 \frac{1}{16\pi^2}\int\frac{\exp[\Delta_1^2-\Delta_2^2]{\rm
 erf}\left[\Delta_1,\sqrt{2}\Delta_1\right]-\{\Delta_1\leftrightarrow\Delta_2\}}{\cos\varphi_1-\cos\varphi_2}
 \cos\nu(\varphi_1+\varphi_2) D\varphi_k,\label{2.3.7}\\
 \hspace*{-0.9cm}\rho_\chi\left(\frac{x}{2n}\right)&=&\int\frac{(-\imath s_1)^{\nu}}{16\pi \widehat{a}^2}
 \frac{e^{-((s_1-x)^2+(s_2+ix)^2)/16\widehat{a}^2}}{s_1-\imath s_2}
 \det\left[\begin{array}{cc} s_1  K_{\nu+1}(s_1) & s_2J_{\nu+1}(
 s_2) \\ K_{\nu}(s_1) & J_{\nu}(s_2) \end{array}\right]\frac{\delta^{(\nu-1)}(s_1)}{(\nu-1)!}d[s].\label{2.3.8}
\end{eqnarray}
 The functions ${\rm sinc}$, ${\rm erf}$, $J_l$, $ K_l$ and $\delta^{(l)}$ are
 the \textit{sinus cardinalis}, the generalized incomplete error function 
 (${\rm erf}(b,c)={\rm erf}(c)-{\rm erf}(b)$), 
 Bessel function of the first kind, the modified one of the second kind and the $l$-th derivative of the
 $\delta$ function, respectively. We employed the abbreviations $D\varphi_k =
 \sin^2((\varphi_1 -\varphi_2)/2)d\varphi_1d \varphi_2$ and 
 $\Delta_j=2\widehat{a}\left(\cos\varphi_j-{x}/{8\widehat{a}^2}\right)$. Notice that only the singular
 part of $K_l$ contributes due to the $\delta$ function and that $\rho_\chi$ vanishes for $\nu=0$.

 For sufficiently small $\widehat{a}$, $\rho_c$ is broadened by a Gaussian along the imaginary axis, the
 oscillations of the continuum limit are distinct and near the real axis it behaves like $y^{\nu+1}$.
 When increasing the lattice spacing the oscillations disappear and the behavior near the real axis
 becomes independent of $\nu$ and $y$. The distribution $\rho_\chi$ develops a box-like shape along
 the imaginary axis with support $|x|\leq8\widehat{a}^2$.  In Fig.~\ref{fig1} we show 3D-plots of $\rho_{\rm c}$.

\begin{figure}
 \includegraphics[width=0.52\textwidth]{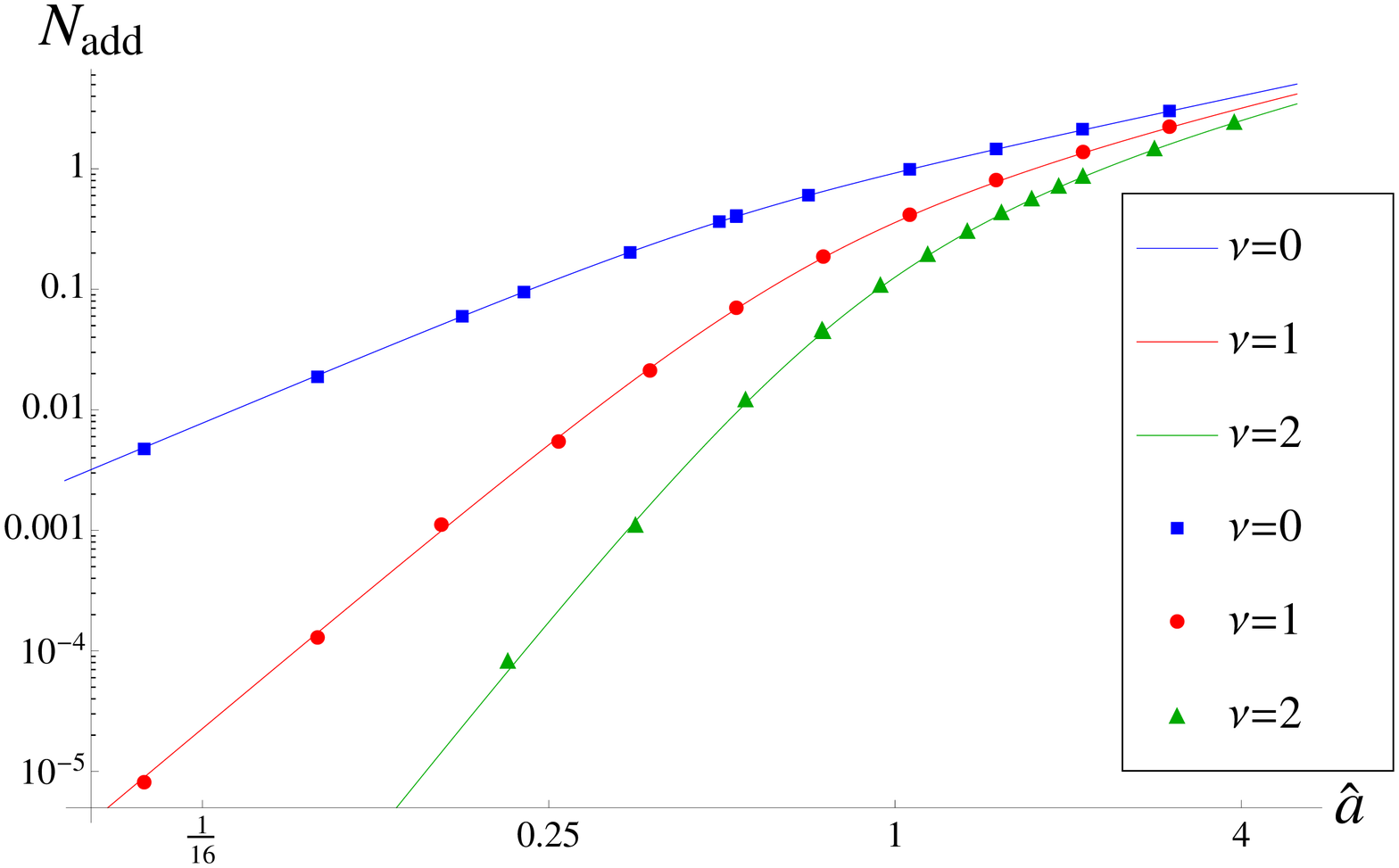}\hfill
 \includegraphics[width=0.48\textwidth]{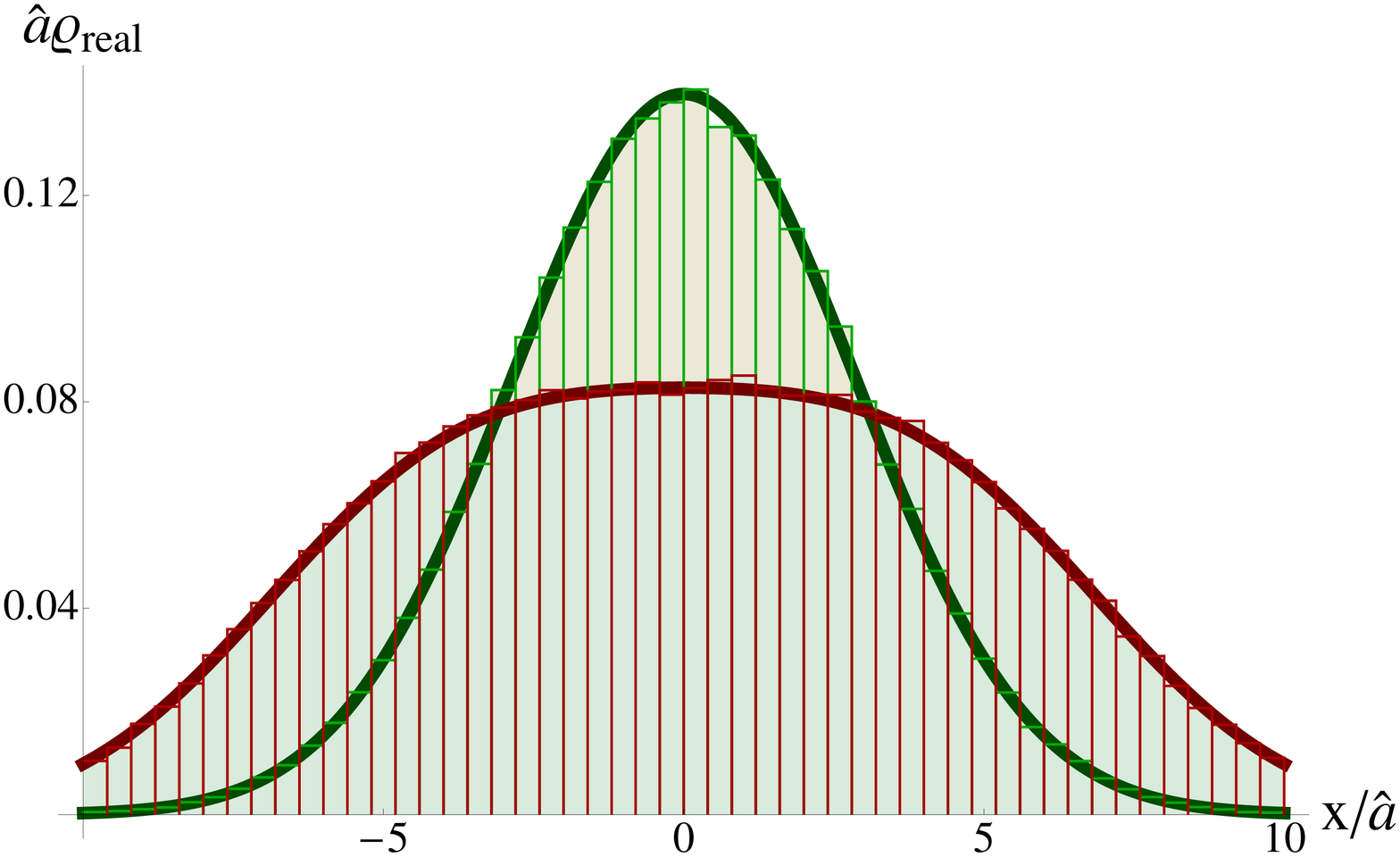}\label{fig2}
 \caption{\textbf{Left:} Log-log-plot of the average number of additional real modes over $\widehat{a}$. The Monte Carlo simulations (symbols) confirm our analytical results (solid curves). The
 matrix dimension and the number of matrices vary in the simulations.
 \textbf{Right:} The distribution $\rho_{\rm real}$ is governed by $\rho_\chi$ at small $\widehat{a}$
 (green, $\widehat{a}=1/\sqrt{200}$) but is increasingly dominated by 
$\rho_{\rm r}$ (red, $\widehat{a}=1/\sqrt{2}$). We compare our analytical
 results (solid curves) with Monte Carlo simulations (histograms, 200000 matrices, $n=50$,
 $\nu=1$, bin size is 0.4).}
\end{figure}

 An important quantity to measure the effect of a finite lattice spacing is the average number of the
  additional real modes. It is given by \cite{kim}
\begin{eqnarray}
 \hspace*{-0.5cm}N_{\rm add}=2\int_{\mathbb{R}} \rho_{\rm r}\left(\frac{x}{2n}\right)dx
 =\int_0^{2\pi}\frac{1-e^{-4\widehat{a}^2\sin^2\varphi}I_0(4\widehat{a}^2\sin^2\varphi)}{4\pi\sin^2\varphi}
\cos2\nu\varphi d\varphi\propto\left\{\begin{array}{cl} \widehat{a}^{2(\nu+1)}, & \widehat{a}\ll1,\\ \widehat{a}, & \widehat{a}\gg1.\end{array}\right. \label{2.3.9}
\end{eqnarray}
 At small lattice spacing contributions for non-zero index are suppressed whereas $N_{\rm add}$  for
 large $\widehat{a}$ is independent of the index. This is shown in Fig.~\ref{fig2}.

At small lattice spacing,  the distribution of the real eigenvalues, $\rho_{\rm real}=\rho_{\rm r}+\rho_{\rm l}=2\rho_{\rm
 r}+\rho_{\chi}$, is approximated  by the $\nu$-dimensional Gaussian unitary ensemble. 
The width is $2\widehat{a}$ in this regime. For
 increasing lattice spacing 
the support increases to $|x|\leq8\widehat{a}^2$ (see Fig. \ref{fig2}) and develops
a square root singularity at the edges, i.e. $\rho_{\rm
 real}(x)=1/[(2\pi)^{3/2}\widehat{a}]+\nu/[\pi\sqrt{(8\widehat{a}^2)^2-x^2}]$. The first term is due to
 $\rho_{\rm r}$ while the singularity comes from $\rho_\chi$. 

\section{Staggered fermions}\label{sec3}
 
In this section we introduce an ensemble to study the taste breaking of the staggered
Dirac operator. For simplicity we only consider the case of two tastes with the ensemble 
given by
\begin{equation}
 \hspace*{-0.1cm}D_{\rm st}=\left(\begin{array}{cccc} 0 & V \\ -V^\dagger &0 \end{array}\right),\
 V=\left(\begin{array}{cc} W+a_1\widetilde{W} & a_2C \\ a_2B & W^{\dagger}-a_1\widetilde{W}^{\dagger}
 \end{array}\right),\ P(D_{\rm st})\propto e^{-n(\tr A^2/2+\tr B^2/2+\tr WW^\dagger +\tr
 \widetilde{W}\widetilde{W}^\dagger)},\label{3.1}
\end{equation}
 where $W,\widetilde{W}$ are $n\times(n+\nu)$ and $B,C$ are $(n+\nu)\times(n+\nu)$ and $(n\times n)$
 complex matrices respectively. The matrix dimension $n$ is identified with the spacetime volume $V$ as
 in the case of the Wilson Dirac operator. 
For $a=0$ this model has two flavors with $\nu$ zero modes for each flavor while at $a\neq 0$ with
 interacting  tastes, the  zero modes are absent and the continuum ${\rm SU}(2)$ flavor
 symmetry is broken.
A more general random matrix model with four tastes and additional
taste breaking terms was introduced in Ref. \cite{Osb10}.

Studying the ensemble~(\ref{3.1}) in the limit, $n\to\infty$, at rescaled quark masses $\widehat{M}=2nM$
and lattice spacing $\widehat{a}_j^2=na_j^2/2$ allows
us to obtain universal results for the  eigenvalue correlations which will be worked out elsewhere.
Here, we evaluate
the partition function of $N_{\rm f}$ fermions corresponding to the matrix model~(\ref{3.1})  which is given by the unitary matrix integral
 \begin{eqnarray}
    \hspace*{-0.3cm}\textrm{Z}&\propto & \int_{{\rm U}(N_{\rm f})}dU
{\det}^\nu U 
\exp\left[\widehat{a}_1^2 \tr[\tau_3U\tau_3U^{\dagger}]+\widehat{a}_2^2\tr
 [(\tau_1U)^2+(\tau_2U)^2+\textrm{cc.}]
 -\tr\widehat{M}(U+U^{\dagger})\right].\label{3.3}
 \end{eqnarray}
 It agrees with the $\varepsilon$  limit 
 of the staggered chiral Lagrangian 
 corresponding to the taste breaking pattern of the
 ensemble~(\ref{3.1}), and is a special case of the result derived in Ref.~\cite{Osb10}.
 
\begin{figure}
\includegraphics[width=0.49\textwidth]{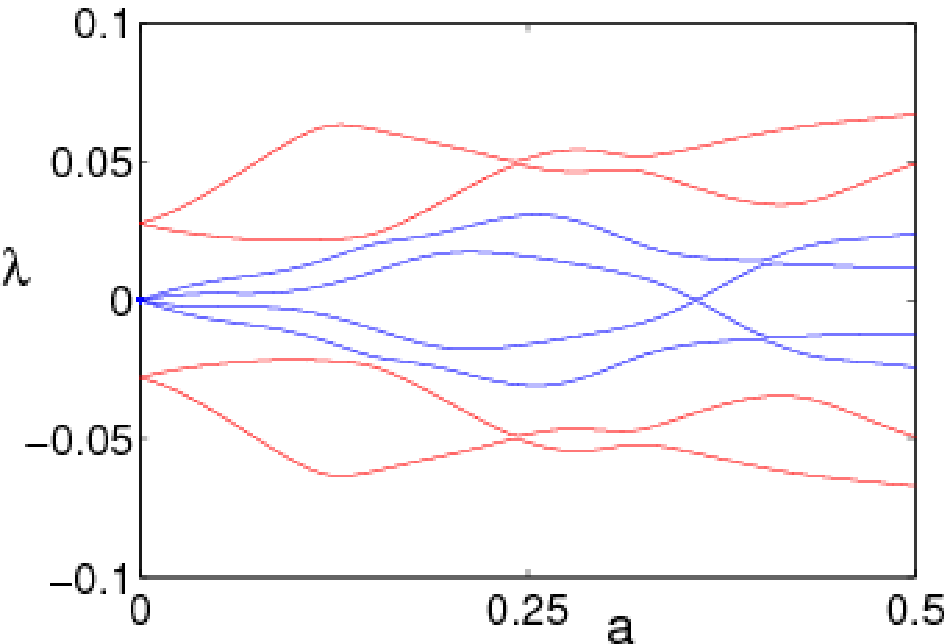}
\includegraphics[height=5cm,width=0.49\textwidth]{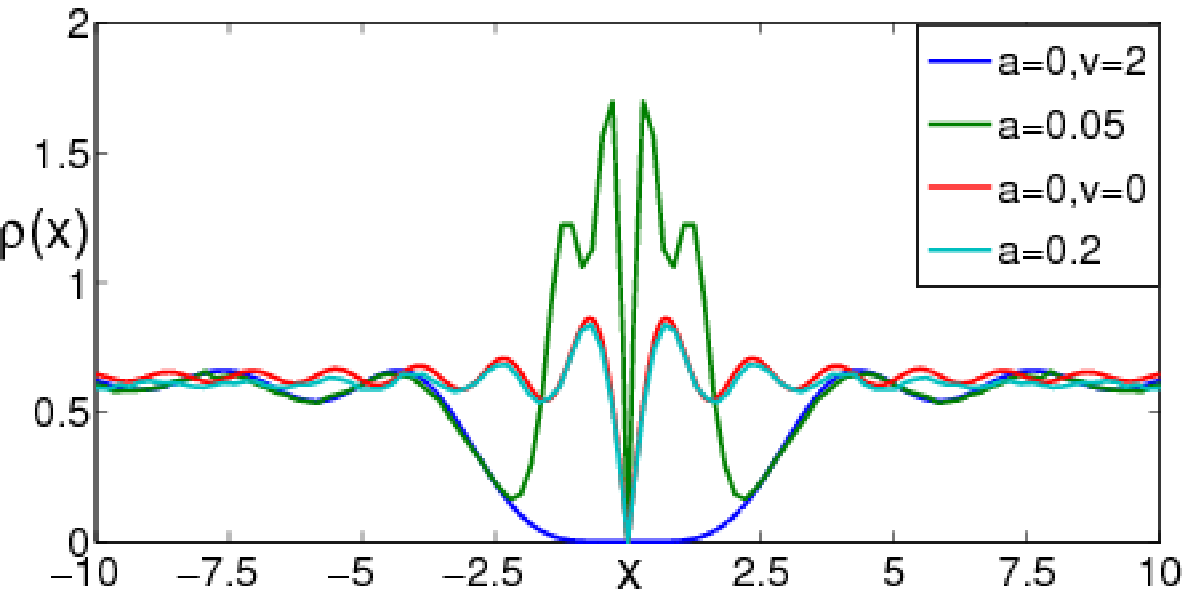}\label{fig3}
\caption{\textbf{Left:} The spectral flow of staggered eigenvalues for $\nu=2$ 
as a function of $a$ using a fixed realization of the random matrix ensemble~(3.1) for $n=50$. The first avoided level
 crossing determines the perturbative regime. \textbf{Right:}
 The quenched spectral densities of $D_{\rm st}$ for values of $a$ (nonrescaled) and $\nu$ 
given in the legend of the figure 
obtained by Monte Carlo simulations with $n=50$ and $100000$ matrices. In both figures we set $a_1=a_2=a$.}
\end{figure}

 The spectral flow as a function of $a$ shows avoided level crossings  
 when lattice artifacts start
 dominating the Dirac spectrum (see Fig.~\ref{fig3} (left)). 
 For small $a$ we observe a 
 linear behavior that follows from perturbation theory. 
 At $a=0$ all eigenvalues have a
 degeneracy of two. The matrix $\widetilde{W}$ in Eq.~(\ref{3.1}) lifts this degeneracy but does not give
 repulsion between the two subspectra. On the other hand, the matrices $B$ and $C$ lift the
 degeneracy and cause a repulsion between the subspectra.

 For $\widehat{a}_j\ll 1$ the spectral density exhibits a chGUE with twice the number of flavors as  for $\widehat{a}_j\gg 1$. 
 At $a_2=0$ the distribution of the zero modes is given by a Dirac delta function with
 weight $2\nu$ at zero. For increasing $a_2$ the distribution of these modes broadens
 gradually. Because they repel each other as well as the other modes, we always find a vanishing
 eigenvalue density at zero.

\section{Conclusions}\label{sec4}

 For the Wilson-Dirac operator the discretization effects become strong when $\widehat{a}\gg0.5$, i.e.
 $ \widetilde{a} \gg 1/(2\sqrt{W_8 V})$. Then  characteristics of the continuum limit like
 the oscillations  in the spectral density or its  behavior near the real axis disappear. In the limit of large $\widehat a$,
 the eigenvalue distributions become
 independent of the index $\nu$, have a support in $|x|\leq8W_8V\widetilde{a}^2$ and the distribution
 for the complex as well as for the eigenvalues of the right handed modes develop plateaus. The chirality
 distribution for large lattice spacing has square root singularities at the edges.

 At sufficiently small lattice spacings the low energy constant $W_8$ can be extracted by combining the
 width of the Gaussian which broadens $\rho_c$ in the $x$ direction, $\sigma = 2 \widetilde{a} \sqrt
 {W_8/V} /\Sigma$, and the spacing of the projected eigenvalues onto the imaginary axis, $ \Delta
 \lambda = \pi /\Sigma V$, yielding $\sigma/\Delta \lambda =2\widehat a /\pi $. Furthermore, additional
 real modes are highly suppressed by non-zero index $\nu$ and, thus, are not much of a problem
 in lattice QCD since most configurations have an index $|\nu|  >  0$ for large volumes.

 The matrix model proposed above is the version with two tastes of the one in Ref.~\cite{Osb10} and describes
 the main features of  the low-lying spectrum of the two dimensional staggered Dirac operator.
 The scale $1/ \sqrt{n}$ (i.e. $1/ \sqrt{V}$ ) determines the appearance of lattice artifacts with
 increasing lattice spacing $a$.
 We expect that it is analytically solvable like the ensemble for the Wilson Dirac
 operator.

\paragraph{Acknowledgements.}

We thank Kim Splittorff for simplifying Eq.~(\ref{2.3.9}) and other helpful comments. MK acknowledges financial support by the Alexander-von-Humboldt Foundation. JV and SZ acknowledge support by
U.S. DOE Grant No. DE-FG-88ER40388.

\end{document}